\documentclass[manuscript=article]{achemso}
\usepackage[version=3]{mhchem} % Formula subscripts using \ce{}
\usepackage{siunitx}
\usepackage{tabularx}
\usepackage{caption}
\usepackage{hyperref}

\usepackage{multicol}
\usepackage{xcolor}
\usepackage{setspace}
\hypersetup{colorlinks,linkcolor={blue},citecolor={blue},urlcolor={blue}}  

\newcommand{\update}[1]{\textcolor{black}{#1}}

\author{\small{Subhadeep Dasgupta}}
\affiliation{\small{Department of Physics, Indian Institute of Science, Bangalore, 560012, India}}
\author{\small{Arun K.S.}}
\affiliation{\small{Department of Physics, Indian Institute of Science, Bangalore, 560012, India}}
\author{\small{K. Ganapathy Ayappa}}
\affiliation{Department of Chemical Engineering, Indian Institute of Science, Bangalore, 560012, India}
\author{\small{Prabal K. Maiti}}
\email{maiti@iisc.ac.in}
\affiliation{\small{Department of Physics, Indian Institute of Science, Bangalore, 560012, India}}
\title{\singlespacing Trajectory Extending Kinetic Monte Carlo Simulations to Evaluate Pure and Gas Mixture Diffusivities through a Dense Polymeric Membrane
  \footnote{Preprint | The Journal of Physical Chemistry B | Accepted on October 25, 2023}
}

\abbreviations{MD, kMC, TEKMC, GCMC, CMS}

\begin{document}
\singlespacing

%\begin{tocentry}
%\centering

%\end{tocentry}

\begin{abstract}

With renewed interest in \ce{CO2} separations, carbon molecular sieving (CMS) membrane performance evaluation requires diffusion coefficients as inputs to have reliable estimate of the permeability. An optimal material is desired to have both high selectivity and permeability. Gases diffusing through dense, CMS and polymeric membranes experience extended sub-diffusive regimes which hinders reliable extraction of diffusion coefficients from mean squared displacement data.
We improve the sampling of the diffusive landscape by implementing the trajectory extending kinetic Monte Carlo (TEKMC) technique to  efficiently extend MD trajectories from ns to $\mu$s timescales. The obtained  self-diffusion coefficient of pure \ce{CO2} in CMS membranes derived from 6FDA/BPDA-DAM precursor polymer melt
is found to linearly increase from $0.8 - \SI{1.3e-6}{\centi\meter\squared\per\second}$ in the pressure range of $1 - \SI{20}{bar}$ which supports previous experimental findings.
We also extend the TEKMC algorithm to evaluate the  mixture diffusivities in binary mixtures to determine the permselectivity of \ce{CO2} in \ce{CH4} and \ce{N2} mixtures. The mixture diffusion coefficient of \ce{CO2} ranges from $1.3 - \SI{7e-6}{\centi\meter\squared\per\second}$ in binary mixture \ce{CO2}:\ce{CH4} which is significantly higher than the pure gas diffusion coefficient.
Robeson plot comparisons show that the  permselectivity obtained from pure gas diffusion data are significantly lower than that predicted using mixture diffusivity data.
Specifically in the case of the \ce{CO2}:\ce{N2} mixture we find that using mixture diffusivities led to permeslectivites lying above the Robeson limit highlighting the importance of using mixture diffusivity data for an accurate evaluation of the membrane performance.
Combined with gas solubilities obtained from grand-canonical Monte Carlo simulations, our work shows that simulations with the TEKMC method can be used to reliably evaluate the performance of materials for gas separations.  
\end{abstract}

\section{Introduction}
\label{sec:introduction}
%%%%%%%%%%%%%%%%%%%%%%%%%%%%%%%%%%%%%%%%%%%%%%%%%%%%
%% Importance of membrane technology
%%%%%%%%%%%%%%%%%%%%%%%%%%%%%%%%%%%%%%%%%%%%%%%%%%%%

Separating \ce{CO2} from natural gas streams primarily involves the separation of \ce{CO2} from gas mixtures.
Separation technologies largely comprise solvent-based extraction techniques, cryogenic distillation, and membrane-based separations~\cite{Carta2015}.
Solvent based extraction techniques, like amine filtration, exhibit long term environmental concerns regarding energy consumption, operating challenges, and emission of hazardous by-products~\cite{nielsen2012atmospheric}.
Currently, industrial applications are dominated by cryogenic distillation.
However, it has serious drawbacks pertaining to high energy requirements and high operating costs~\cite{BHATTA2015171}.
Membrane-based gas separation offers significant advantages, due to increased stability, scalability, with potential to improve performance based on synthesis of novel materials, while being relatively hazard-free~\cite{koros2004evolving}.
Carbon based membranes have shown a lot of promise for effectively separating \ce{CO2} from natural gas streams. 
Carbon molecular sieving (CMS) membranes are carbon-based high-performance gas separation membranes, derived from the pyrolysis of polymeric precursors in vacuum or inert atmosphere.
CMS membranes present a bimodal pore-size distribution and a rich network of interconnected micropores $(\sim 0.7-\SI{2}{\nano\meter})$ and ultra-micropores $(< \SI{0.7}{\nano\meter})$ that can differentiate between pairs of gas molecule having similar kinetic diameters~\cite{kamath2020pyrolysis,Vu_Miller_IECR_2002,Kumar_Koros_JMemSci_2019,rungta2017carbon}.
Similar pore network topologies are observed in a variety of other polymeric membranes, enabling one to modify membranes by varying the chemistry and processing conditions~\cite{SONG2c01450}.
Membrane based technologies also have the advantage of excellent gas selectivity with high gas permeability, chemical and thermal stability~\cite{li2018review, williams2006analysis}.
CMS membranes show great promise in carbon capture and storage applications, with their monomer composition and choice of precursor playing an important role in determining their gas adsorption and selectivity performances.

%%%%%%%%%%%%%%%%%%%%%%%%%%%%%%%%%%%%%%%%%%%%%%%%%%%%
%% Challenges
%%%%%%%%%%%%%%%%%%%%%%%%%%%%%%%%%%%%%%%%%%%%%%%%%%%%
Molecular simulations are widely used to study gas adsorption and separation processes.
Grand-canonical Monte-Carlo (GCMC) simulations have been used to obtain adsorption isotherms for a variety of molecules and microporous materials such as metal organic frameworks (MOFs), zeolites and carbon based materials.
These simulations are carried out under equilibrium conditions.
In contrast the transport mechanism of gases inside membranes occur in non-equilibrium conditions across an applied pressure gradient between the upstream feed gas to be separated and the downstream gas.
In this regard, polymeric membranes have been widely studied for effective \ce{CO2} separations.
Gas transport through dense polymer membranes are generally modelled by the sorption-diffusion mechanism~\cite{paul1976solution, wijmans1995solution, lonsdale1982growth}.
Gas molecules first adsorb in the membrane at the upstream conditions, diffuse under the influence of a chemical potential gradient, and  finally desorb from the membrane at the downstream compartment.
The diffusive flux of the gas molecule through the membrane is measured using its permeability,
\begin{equation} \label{eq 1.1} 
 \mathcal{P} = SD     
\end{equation}
where $S$ and $D$ are the solubility and diffusion coefficients of the gas molecule respectively in the membrane~\cite{suloff2002permeability}.
High permeability values can be achieved by increasing either \update{$D$ or $S$ or both $D$ and $S$}.
If the upstream pressure is much larger than the downstream pressure then the solubility $S$ can be obtained from an equilibrium GCMC simulation evaluated at the thermodynamic conditions at the upstream pressure.
Molecular dynamics (MD) simulations have been widely used  to evaluate the diffusion coefficient, $D$ in a variety of microporous materials~\cite{anastasios2006, jacs.5b08746,kohen101021,PhysRevE.65.061202,kumar2017porphyrin}.
Due to the complex porous networks present in glassy polymeric membranes and their derivatives, there is a drastic slowing down of the dynamics of the diffusing species.
As a result carrying out all-atom MD simulations to adequately sample
the diffusive regime and reliably evaluate $D$ is a computational challenge. 
Alternate approaches have been devised to overcome this limitation with the primary goal of evaluating the self-diffusion coefficients of slowly diffusing gas species. 
Thornton \textit{et al.} presented an approximate dependence of gas diffusivity on fractional free volume of the porous media and the kinetic diameter of the gas, derived empirically from available experimental data~\cite{Thornton2009}.
Bousige \textit{et. al.} analyzed the residence and relocation times of fluid in ultraconfining disordered porous materials by mapping MD simulations to mesoscopic random walks~\cite{bousige2021bridging}.
Neyertz \textit{et al.} proposed a variant of kinetic Monte Carlo (kMC), referred to as trajectory extending kinetic Monte Carlo (TEKMC)~\cite{neyertz2010trajectory}.
The algorithm has been successfully used to calculate the diffusion coefficients of small molecules such as H$_2$O, O$_2$, \ce{N2} in glassy polymers ~\cite{neyertz2010trajectory, neyertz2010carbon}.

While assessing a given material for a particular gas separation technology both solubility and  diffusion coefficients in a gas mixture need to be evaluated.
The gas solubility can be obtained from mixture GCMC simulations, however evaluating diffusion coefficients of mixtures is more challenging~\cite{ROBESON1991165,freeman1999basis}.
There are several reports of computing gas permeabilities in polymers, metal organic frameworks, zeolites using both pure and mixture gas data while comparing with the Robeson upper limit ~\cite{robeson2008upper,ROBESON2009178} where higher selectivity is correlated with lowered permeability.  
These plots are used as a standard to assess the performance of newly fabricated adsorbents where the goal is to design a material with both high selectivity and permeability. 
There are also ongoing efforts to redefine and increase the upper limits for different materials~\cite{C9EE01384A,YANG2020631,HAN2021119244}.
With a wide range of data available, machine learning based techniques have also been leveraged to understand and find materials that can surpass the Robeson upper bounds~\cite{sanat_sciadv.aaz4301}. 
Recent attempts to address mixture diffusivities from MD simulations showed that gas molecules tend to oscillate between adsorption sites~\cite{li2022molecular} making it difficult to evaluate the long-time gas diffusivity. Since the evaluation of diffusion coefficients in mixtures is challenging several studies evaluate the permeabilities using pure component diffusion coefficients.
Semi-empirical approaches which have been developed with varying degrees of success to obtain mixture diffusion coefficients from pure diffusivity data~\cite{KRISHNA2000477}.   
Lattice-based techniques have been used to compute diffusion of gases inside such variable pore structures~\cite{Sameer2013, Mahmood2015174}, however they require various parameters which are difficult to determine.
In this work, to overcome these limitations, we extend the TEKMC technique to reliably obtain mixture diffusivities of gases inside the 6F-CMSM membrane. Robeson plot comparisons are made with permeability data obtained from both pure and mixture diffusivities to highlight the limitations of using pure component diffusivity data while designing membranes for gas separation applications.  

In our previous study~\cite{DASGUPTA2022121044}, we modelled CMS membranes derived from 6FDA/BPDA-DAM precursor polymer melt ~\cite{roy2020investigations, kiyono2010effect} (denoted as 6F-CMSM).
The molecular structures were built in close correspondence with experimental compositions using density functional theory (DFT) for optimizing the two monomeric units (pyridine and pyrrole) followed by their polymerization using all-atom MD simulations.
GCMC simulations were used to obtain gas adsorption isotherms inside the membrane for a range of pressures. Our simulated 6F-CMSM morphologies helped understand the importance of the length of carbon chains in determining density of the membrane and their accessible pore volumes, which in turn affect their adsorption performances.
More details about the CMS membrane simulations can be found in our previous publication~\cite{DASGUPTA2022121044}.
In continuation of our previous study, here we analyze the dynamics of the adsorbed gases through the porous networks of the 6F-CMSM.
The solubility coefficients needed in this work are obtained from the previous gas adsorption isotherms under different pressures.
Predicting accurate values of $\mathcal{P}$ requires estimating  both $S$ and $D$ for each gas component.
The novelty of our work lies in using TEKMC to obtain self-diffusion coefficients of \ce{CO2}, \ce{CH4} and \ce{N2} in pure gases and the multicomponent diffusivities in binary mixtures, as a function of pressure.
We obtain the permeability for pure gases and binary \ce{CO2}:\ce{CH4} and \ce{CO2}:\ce{N2} mixtures to evaluate the membrane performance.
The manuscript is organized as follows.
We first discuss in detail the methodology used to perform the TEKMC steps.
Next we study the variation of gas diffusivities and ultimately their permeation through the 6F-CMSM.
We study the relative permeabilities of these gases in both the pure state and binary mixtures in the 6F-CMSM polymeric membrane to understand its dependence on system pressure and gas compositions.
We finally present the Robeson plots for the different systems to illustrate the importance of using accurate  mixture diffusion coefficients while assessing a materials capacity for separating a given gas mixture.  

\section{Method}
\label{sec:modelling}
\begin{figure}[ht]
    \centering
    \includegraphics[width=\textwidth]{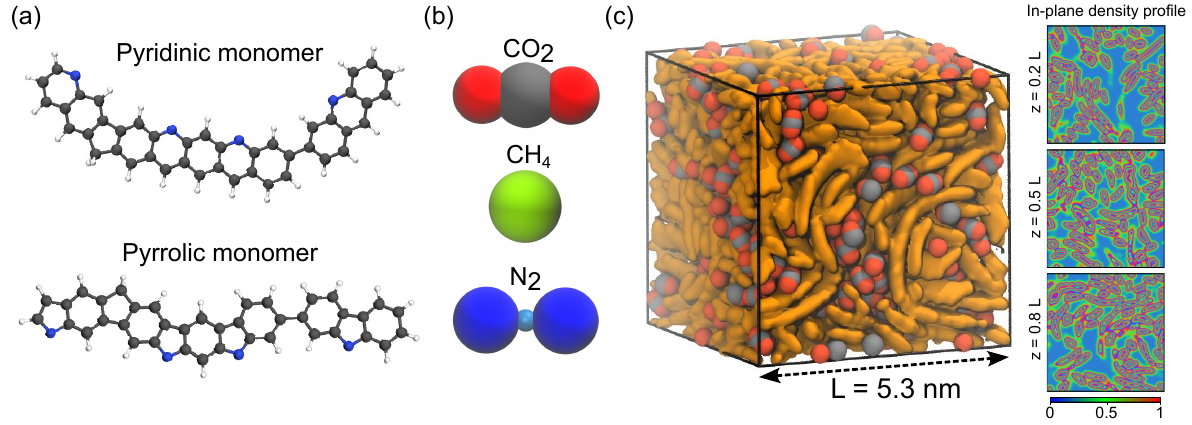}
    \caption{
    (a) Atomic structures of pyridinic and pyrrolic monomer fragments that cross-link and polymerize to form the derived 6F-CMS membrane.
    Gray denotes carbon, blue denotes nitrogen, and white denote hydrogen atoms
    (b) Atomic structure for the three gases considered in this work based on the TraPPE force field~\cite{trappe}.
    Three-site \ce{CO2} is modelled with two oxygen (red) and one carbon (gray) atoms.
    \ce{CH4} is represented with united atom model, having a single bead (green).
    The three site \ce{N2} with two nitrogen atoms (blue) and one fictitious centre of mass bead (cyan).
    (c) Snapshot showing the derived 6F-CMSM containing adsorbed \ce{CO2} in the pores, corresponding to $T = \SI{308}{\kelvin}$ and $P = \SI{20}{bar}$.
    The surface of the cross-linked monomers are colored orange.
    Shown in the right are in-plane density profiles for three cross-sections.
    The blue regions represent the pores and green to red regions denote presence of 6F-CMSM atoms.
    The difference in porous structure across the z-plane indicate formation complex network pathways along the bulk of the 6F-CMSM.
    These interconnected pores govern the dynamics and diffusion of gases inside the 6F-CMSM.
    }
    \label{fig:membrane}
\end{figure}
%%%%%%%%%%%%%%%%%%%%%%%%%%%%%%%%%%%%%%%%%%%%%%%%%%%%
%% CMS details
%%%%%%%%%%%%%%%%%%%%%%%%%%%%%%%%%%%%%%%%%%%%%%%%%%%%
\subsection{Computational modeling and simulation}
The 6F-CMSM membrane used in this work is composed from pyridinic and pyrrolic monomer units shown in Fig.~\ref{fig:membrane}a.
These monomers polymerize to form various morphologies, among which we select the structure with closest resemblance to experimental gas adsorption isotherms as discussed in our earlier work~\cite{DASGUPTA2022121044}.
The structure of \ce{CO2}, \ce{CH4} and \ce{N2} molecules used are shown in Fig.~\ref{fig:membrane}b.
The interactions involving the gas molecules are modeled using the TraPPE force fields~\cite{trappe} and the 6F-CMSM atoms are modelled using modified Dreiding~\cite{roy2020investigations, mayo1990dreiding}.
\update{The force field values used in this work are reported in Table~\ref{table:parameters} of the Appendix.}
The distribution of gas molecules inside the 6F-CMSM are obtained from GCMC simulations as discussed in our earlier work.
\update{To account for the structural rearrangement of monomer constituents upon interaction with gas molecules, the 6F-CMSM is modelled as a flexible membrane.}
We first perform all-atom MD simulations of 6F-CMSMs along with the adsorbed gases for \SI{20}{\nano\second} duration in an NVT ensemble using LAMMPS~\cite{plimpton2007lammps}.
The temperature of the system is maintained at $\SI{308}{K}$ using a Nosé-Hoover thermostat.
The coordinates of the gas molecules from the MD simulation are stored at an interval of $\tau_{MD} = \SI{10}{\pico\second}$.
Fig.~\ref{fig:membrane}c shows a snapshot from our all-atom MD simulation of 6F-CMSM loaded with \ce{CO2} corresponding to $\SI{20}{bar}$ pressure.
The porous nature of the membrane for different cross-sections along the bulk of the membrane is also visible from the simulation snapshot.

\subsection{Implementing kinetic Monte Carlo}
The simulation box is uniformly divided into voxels of grid size $(d_{grid})$ along $x, y,$ and $z$ directions. Each voxel is identified by a single index, spanning from $1$ to $n_{voxel} = (L/d_{grid})^3$, for a cubic simulation box having length $L$ on each side.
All gas molecules are mapped from their real-space $(x,y,z)$ coordinates to their respective voxel indices. The trajectories of the gas molecules from the MD simulations are used to compute the transition probability matrix $(\pi_{ij})$ required as input for the TEKMC simulation.
We define the probability of transition between voxels $i$ and $j$, given that the gas is in voxel $i$ as,
\begin{equation}\label{eq 4.1}
\pi_{ij} = \frac{N_{ij}}{N_i}
\end{equation}
where $N_{ij}$ is the number of transitions between voxels $i$ and $j$ and $N_i$ is the number of times the gas molecule visits voxel $i$.
Since the transition probability matrix is constructed from the all-atom MD trajectories, $\pi_{ij}$ implicitly takes into account all the gas-gas and gas-membrane interactions. 
The probability matrix is symmetrized to maintain detailed balance, such that $\pi_{ij}=\pi_{ji}$, which is equivalent to
\begin{equation}
{N_{ij} = N_{ji} = (N_{ij}^{'} + N_{ji}^{'})/2}
\end{equation}
where $N_{ij}^{'}$ and $N_{ji}^{'}$ are the recorded number of transitions from voxel $i$ to $j$ and $j$ to $i$ respectively.
Voxels with $\pi_{ii} = 0$ indicate regions inside the simulation box where the atoms of the CMS molecules are located.
The values of $(\pi_{ii})$ for \ce{CO2}, \ce{CH4} and \ce{N2} for $\SI{1}{\bar}$ and $\SI{20}{\bar}$ pressures are shown in Fig.~\ref{fig:probability} of the appendix.
In case of binary gas mixtures, $\pi$ of each component is obtained from the trajectory of that component.
The interaction between two components changes the probability distribution from the pure gas system, shown in Fig.~\ref{fig:mix_probability} of the appendix for $\SI{20}{\bar}$ pressure.
From conservation of number of particles, we have the condition $\Sigma_{j} \pi_{ij} = 1, \forall i$.
For each kMC move, a walker is placed randomly in one of the visited voxels $i$ such that $\pi_{ii} \neq 0$.
The voxel $j$ that the walker visits form $i$ is determined from the sum, $\sigma(j') = \Sigma_{1}^{j'}\pi_{ij}$.
The destination voxel is identified to be $j = j'$ when $\sigma(j') \geq r$ is just satisfied, where $r$ is a random number drawn from a uniform distribution.
A total of 5000 random walkers are inserted for each simulation and the time associated with each kMC step is $\tau_{kMC}$.
The trajectories of these random walkers are mapped back from the voxel index to the real space coordinates and their average mean squared displacement (MSD) is determined. 

%%%%%%%%%%%%%%%%%%%%%%%%%%%%%%%%%%%%%%%%%%%%%%%%%%%%
% Parameters
%%%%%%%%%%%%%%%%%%%%%%%%%%%%%%%%%%%%%%%%%%%%%%%%%%%%
\begin{figure}[t!]
    \centering
    \includegraphics[height=5cm]{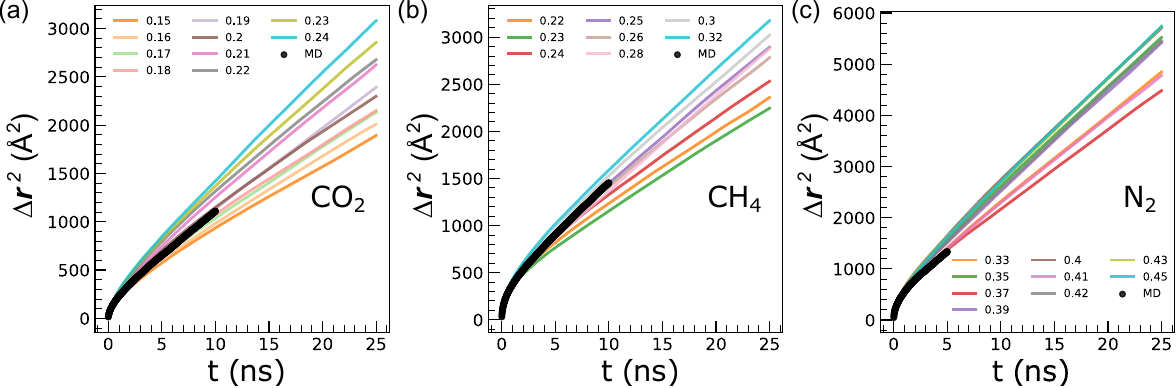}
    \caption{
    Mean squared displacement $(\Delta r^2)$ of (a) \ce{CO2}, (b) \ce{CH4} and (c) \ce{N2}, inside 6F-CMSM at a temperature of $\SI{308}{\kelvin}$ and $\SI{20}{bar}$ pressure.
    The thick black line shows the values obtained from the all-atom MD simulations.
    The finer colored lines show the values obtained from \SI{50}{\nano\second} TEKMC simulations for different values of $d_{grid}$ (nm).
    The values of $d_{grid}$ are tuned to obtain an MSD having least mean squared error \textit{w.r.t.} the values obtained from MD simulation.
    }
    \label{fig:optimization}
\end{figure}

\begin{table}[b!]
    \caption{Optimum values of $d_{grid}$ (nm) corresponding to lowest MSE between MSD obtained from TEKMC and MD simulations for the three gases.}
    \label{table:d_grid}
    \footnotesize
    \begin{tabular}{cccccccc}
    \hline
P (bar)   & 0.1  & 0.6  & 1    & 6    & 10   & 15   & 20    \\
    \hline
    \hline
\ce{CO2} & 0.41 & 0.32 & 0.22 & 0.19 & 0.20  & 0.20  & 0.20   \\
\ce{CH4} & 0.61 & 0.42 & 0.34 & 0.30  & 0.32 & 0.24 & 0.25  \\
\ce{N2}  & 0.60  & 0.55 & 0.47 & 0.45 & 0.44 & 0.35 & 0.33 \\
    \hline
\end{tabular}
\end{table}

\subsection{Optimization to obtain diffusion and permeability}
The optimal values of $\tau_{MD}$ and $d_{grid}$ combinations first needs to be determined for obtaining the correct MSD from the TEKMC simulation.
Different techniques can been used to obtain the time interval between two successive kMC steps, $\tau_{kMC}$ ~\cite{tien2018generic}.
We follow the procedure prescribed by Neyertz \textit{et al.} where $\tau_{kMC} = \tau_{MD}$ $(= \tau)$.
Fig.~\ref{fig:optimization} illustrates the values of MSD obtained from all-atom MD and TEKMC simulations for \ce{CO2}, \ce{CH4} and \ce{N2} at $\SI{20}{bar}$ pressure.
For a diffusing particle the MSD, $\langle \Delta \mathbf{r}(t)^2 \rangle \propto t^\alpha$.
The value of the exponent $\alpha$ from the MSD data obtained from  MD simulations lies in the range of $0.6 \sim 0.8$ indicating the sub-diffusive nature of the diffusing molecule in the polymeric membrane.
In the diffusive regime where the value of $\alpha \simeq 1$, the diffusivity is obtained from the Einstein relation,

\begin{equation} \label{eq 1.2}
D = \lim_{t\rightarrow\infty}\frac{1}{6t}\langle\Delta \mathbf{r}(t)^2\rangle = \lim_{t\rightarrow\infty}\frac{1}{6t}\langle[\mathbf{r}(t+t_0) - \mathbf{r}(t_0)]^2 \rangle
\end{equation}
where $\langle ... \rangle$ represent averages over all particles, and shifted time origins ($t_0$).
The TEKMC simulations enable us to access the Brownian regime for restricted diffusing molecules which predominantly sample the sub-diffusive regime during the small timescale of MD simulations. 
In order to determine the optimal value of $d_{grid}$
we used the lowest mean squared error (MSE) between MSDs obtained from the MD simulation and that obtained from $\SI{100}{\nano\second}$ TEKMC simulations.
The optimum values of $d_{grid}$ for the different cases considered are shown in Table~\ref{table:d_grid}.
To determine the diffusion coefficients we carry out additional TEKMC simulations for three different $d_{grid}$ values, which include the optimal $d_{grid}$ (Table~\ref{table:d_grid}) as well as simulations at $d_{grid} \pm \SI{0.01}{\nano\meter}$. 
For these three different values of $d_{grid}$, random walks are performed up to $\SI{5}{\micro\second}$ to ensure all gases attain the Brownian diffusion regime and the reported values of $D$ are an average over these TEKMC simulations. 
%%%%%%%%%%%%%%%%%%%%%%%%%%%%%%%%%%%%%%%%%%%%%%%%%%%%
% Implementation
%%%%%%%%%%%%%%%%%%%%%%%%%%%%%%%%%%%%%%%%%%%%%%%%%%%%
The in-house TEKMC code developed in this study was validated by reproducing the diffusivity of bulk water and is discussed in section \ref{sec:validation} of the appendix.
From TEKMC, we obtain diffusion coefficient of SPC/E water to be \update{$2.47 \pm \SI{0.15e-5}{\centi\meter\squared\per\second}$} at $\SI{300}{\kelvin}$, in agreement to literature~\cite{doi:10.1021/jp003020w}\update{~\cite{ioannis_2019}}.
The algorithm is generic and can be used for other systems where such kMC algorithms are applicable.
The solubility coefficient of the $i^{th}$ gas species $(S_i)$, in a membrane is the ratio between its volumetric loading $V_{L,i}$ and its partial pressure $p_i$ at a constant temperature~\cite{ALEXANDERSTERN19941},
\begin{equation}\label{eq 2.1}
    S_i = \frac{V_{L,i}}{p_i}
\end{equation}
The adsorption isotherm is expressed in terms of the volumetric loading,
\begin{equation}
{V_{L,i}=\frac{N_i(T, P)}{V_{{CMS}}} \frac{k_{\mathrm{B}} T_{S T P}}{P_{S T P}}}
\end{equation}
where $N(T,P)$ is the number of gas molecules adsorbed in the 6F-CMSM membrane at temperature $T$ and pressure $P$, $V_{CMS}$ is the volume of  the 6F-CMSM membrane, $k_B$ is the Boltzmann constant, $T_{STP}$ and $P_{STP}$ correspond to standard temperature and pressure respectively.
The equilibrated values of $V_L$ were obtained using the Peng-Robinson equation of state~\cite{peng1976new} implemented in RASPA~\cite{dubbeldam2016raspa}, detailed in our previous work~\cite{DASGUPTA2022121044}.
The solubility coefficients can be estimated accurately using the above equations, ultimately allowing us to compute their corresponding permeabilities. 

%%%%%%%%%%%%%%%%%%%%%%%%%%%%%%%%%%%%%%%%%%%%%%%%%%%%
% Results
%%%%%%%%%%%%%%%%%%%%%%%%%%%%%%%%%%%%%%%%%%%%%%%%%%%%
%%%%%%%%%%%%%%%%%%%%%%%%%%%%%%%%%%%%%%%%%%%%%%%%%%%%
% Exponent
%%%%%%%%%%%%%%%%%%%%%%%%%%%%%%%%%%%%%%%%%%%%%%%%%%%%
\begin{figure}[t]
    \centering
    \includegraphics[width=0.95\textwidth]{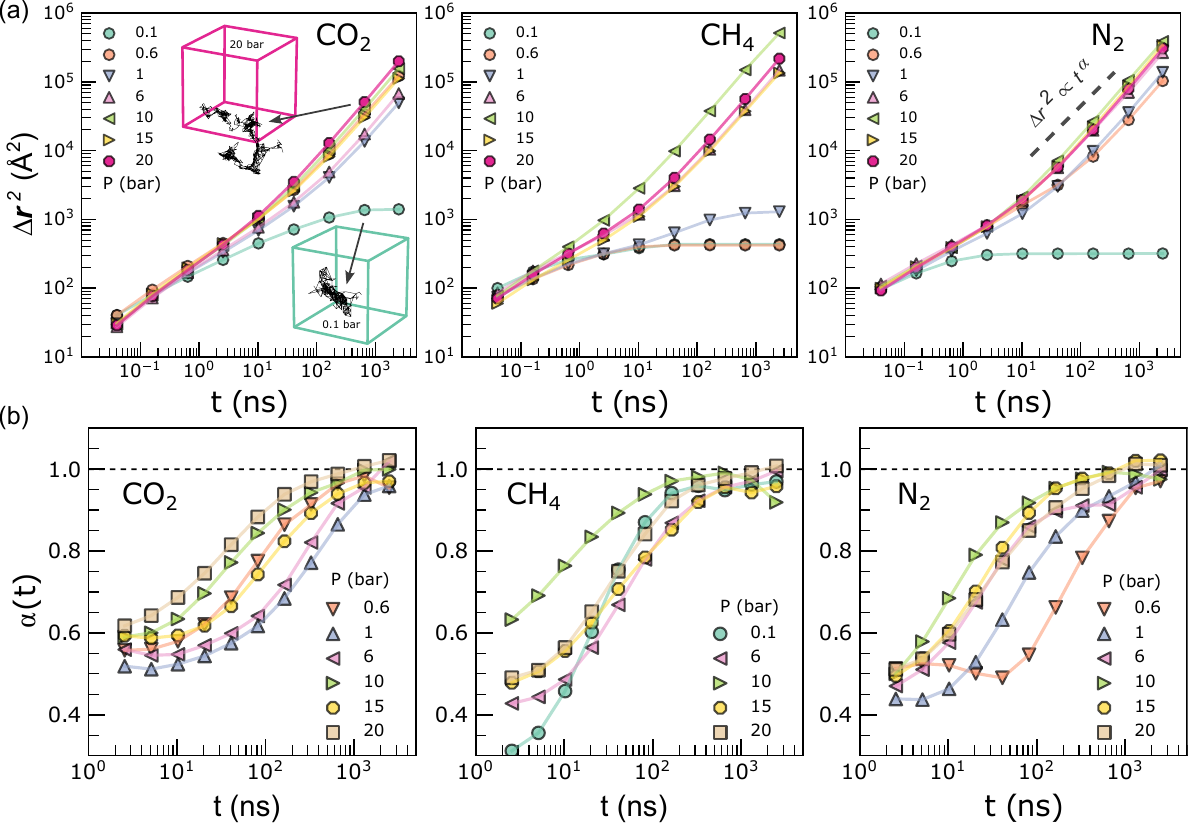}
    \caption{
    (a) MSD $(\Delta \mathbf{r}^2)$ versus time for different gases \ce{CO2}, \ce{CH4}, and \ce{N2}.
    The dynamics of the gas molecules vary significantly depending on the system pressure (indicated by different colors shown in legend).
    The inset shows two sample, unwrapped trajectories of a walker representing \ce{CO2} molecules inside 6F-CMSM at $0.1$ and $\SI{20}{bar}$ pressures obtained from kMC.
    (b) The growth of MSD exponent $(\alpha)$ versus time for the three gases is presented for systems that attain the Brownian regime successfully.
    The dashed line shows the reference level of $\alpha = 1$.
    The time taken to reach $\alpha = 1$ decreases with increasing pressure.
    }
    \label{fig:exponent_series}
\end{figure}

\section{Results and discussion}
\label{sec:results}

\subsection{Gas dynamics inside 6F-CMSM}
Fig.~\ref{fig:exponent_series}a shows the averaged MSD $(\Delta \mathbf{r}^2)$ of \ce{CO2}, \ce{CH4} and \ce{N2} for the entire range of system pressures obtained from $\SI{5}{\micro\second}$ TEKMC simulations and the corresponding values of the exponent $\alpha$ are illustrated in Fig.~\ref{fig:exponent_series}b.
The time taken to attain the Brownian regime for the three gases is significantly higher at low pressures compared to the time taken at higher pressure.
We observe a slow growth of the exponent towards $\alpha = 1$ and the Brownian regime is observed above a sampling time of $\SI{1}{\micro\second}$.
The inset in Fig.~\ref{fig:exponent_series}a illustrates two representative trajectories for \ce{CO2} molecules corresponding to $0.1$ and $\SI{20}{bar}$ pressures.
At low pressures $(\SI{0.1}{bar})$ gas molecules remain trapped in higher energy adsorption sites restricting displacements to a small region of the CMS membrane and results in a sub-diffusive nature of the MSD $(\alpha < 0.7)$ even after $\SI{5}{\micro\second}$.
At higher pressures, a distinct crossover from sub-diffusive to diffusive regimes is observed. 
We note that at higher pressure, increased gas uptake  results in swelling of 6F-CMSM giving rise to flexible permeation pathways~\cite{DASGUPTA2022121044}.

%%%%%%%%%%%%%%%%%%%%%%%%%%%%%%%%%%%%%%%%%%%%%%%%%%%%
%% Analysis
%%%%%%%%%%%%%%%%%%%%%%%%%%%%%%%%%%%%%%%%%%%%%%%%%%%%
\begin{figure}[t!]
    \centering
    \includegraphics[width=\textwidth]{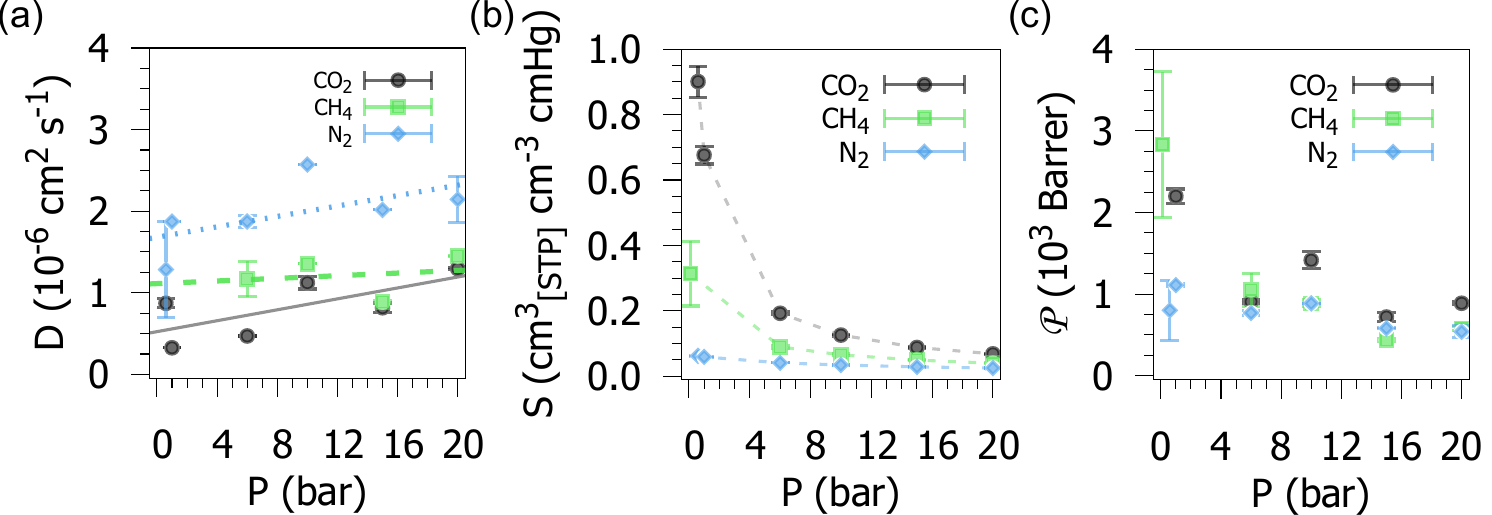}
    \caption{
    Dependence of (a) self-diffusion coefficient, $D$, (b) solubility coefficient, $S$, and (c) permeability, $\mathcal{P}$ of pure \ce{CO2}, \ce{CH4}, \ce{N2} adsorbed inside 6F-CMSM for increasing pressure $P$ at $T = \SI{308}{\kelvin}$.
    The data shown are averaged over independent kMC simulations using three best grid sizes for each pressure point.
    }
    \label{fig:analysis}
\end{figure}

The diffusion coefficients obtained from the Brownian regimes,  $(\alpha \simeq 1)$, for different pressures are shown in Fig.~\ref{fig:analysis}a.
In literature, the experimentally reported value of $D$ for \ce{CO2} inside 6F-CMSM derived from 6FDA/BPDA-DAM pyrolysed at $\SI{550}{\celsius}$ is $\SI{1.32e-6}{\centi\meter\squared\per\second}$~\cite{STEEL20051843}.
Our calculated value of $D_{CO_2}$ is in excellent agreement with the experimental value.
From the values obtained in Fig.~\ref{fig:exponent_series}, the estimated timescale for obtaining diffusive regime at presure $P = \SI{20}{\bar}$ is around $\SI{500}{\nano\second}$ for \ce{CO2}.
For cross-validation against all-atom MD results, we performed an additional all-atom MD simulation for $\SI{500}{\nano\second}$.
The diffusion coefficient from the all-atom MD $\sim \SI{1.23e-6}{\centi\meter\squared\per\second}$.
Both the experimental and all-atom MD values are close to the converged value obtained using TEKMC.
This remarkable agreement further allows us to conclude that the algorithm  can successfully extrapolate short MD results to longer duration and can be successfully used for computing diffusion of confined gas molecules.
We also observe that a sufficient number of percolating pathways are required during the atomistic MD simulation in order to adequately access all possible pores available for the gas molecules to diffuse through.
From the dependence of $D$ on $P$, in Fig.~\ref{fig:analysis}a we observe a linear increase in the diffusivities of all three gases inside the 6F-CMSM as a function of pressure.
The loading-dependent diffusion coefficients have been observed for hydrocarbons in metal organic frameworks~\cite{jacs.5b08746} which lead to structural changes of the adsorbing framework~\cite{ja805129c}.
There is a monotonic increase of $D$ in the pressure range studied for all gases, which is non-trivial and to the best of our knowledge not reported earlier for CMS membranes.
We observe that \ce{N2} diffuses faster than \ce{CO2} inside 6F-CMSM which is similar to the diffusive nature inside zeolites reported in literature~\cite{kohen101021}.

%%%%%%%%%%%%%%%%%%%%%%%%%%%%%%%%%%%%%%%%%%%%%%%%%%%%
%% Solubility
%%%%%%%%%%%%%%%%%%%%%%%%%%%%%%%%%%%%%%%%%%%%%%%%%%%%
\subsection{Solubility and permeability of pure gases}
The solubilities, $(S)$ of \ce{CO2}, \ce{CH4}, and \ce{N2} versus pressure for 6F-CMSM corresponding to temperature $\SI{308}{\kelvin}$ using the values of $V_L$ from our previous study of gas uptake in 6F-CMSM~\cite{DASGUPTA2022121044} and calculated using Eq.~\ref{eq 2.1} are shown in Fig.~\ref{fig:analysis}b.
The values of $S$ for \ce{CO2} are the highest followed by \ce{CH4}.
\ce{N2} solubilities are the lowest and independent of pressure.
We observe a monotonic decrease in the solubility coefficient with increase in pressure for all gases inside the 6F-CMSM.
Permeability $(\mathcal{P})$ is then obtained using Eq.~\ref{eq 1.1} in units of Barrer and shown in Fig.~\ref{fig:analysis}c.
In general, the permeability values for \ce{CO2} are the highest when compared with \ce{CH4} and \ce{N2}.
We also observe that the permeability decreases with pressure with the most noticeable variation seen in \ce{CO2} and \ce{CH4}, the effect is less pronounced for \ce{N2}. 
The linear growth of $D$ with $P$, signifies that any variation in $\mathcal{P}$ is strongly dominated by $S$.

\subsection{Permselectivity in binary gas mixtures}
%%%%%%%%%%%%%%%%%%%%%%%%%%%%%%%%%%%%%%%%%%%%%%%%%%%%
%% Permselectivity
%%%%%%%%%%%%%%%%%%%%%%%%%%%%%%%%%%%%%%%%%%%%%%%%%%%%
In order to understand the relative permeability of different gases inside 6F-CMSM, we also compute the permselectivity $(\alpha_\mathcal{P})$ of \ce{CO2} in binary gas mixtures of \ce{CO2}:\ce{CH4} $(50:50$ and $10:90)$ and \ce{CO2}:\ce{N2} $(20:80)$.
The permselectivity in gas mixture is given by,
\begin{equation}
    \alpha_\mathcal{P} = \frac{\mathcal{P}_{X}}{\mathcal{P}_{Y}}
\end{equation}
where $X$ denotes \ce{CO2} and $Y$ denotes \ce{CH4} or \ce{N2} depending on the composition of gas mixture.
The diffusion coefficient of gas molecules are influenced by their interactions with the  membrane  as well as with neighboring gas molecules.
Consequently, the diffusion coefficient of each component in a mixture deviates from the corresponding pure gas diffusion coefficients.
The deviation is non-trivial and there is an on-going effort to understand multicomponent gas diffusion in different adsorbing frameworks~\cite{anastasios2006, kohen101021,PhysRevE.65.061202, kamala0548321, Ting501488s}.
To address this challenge, we extended the TEKMC simulations to gas mixtures of different compositions.
The probability matrix for the binary gas mixture is created by tracking the positions of individual gas species during the MD runs as was done for the pure single component system  described earlier.  
We performed separate TEKMC simulations for each species in the binary gas mixture system, to obtain the corresponding mixture diffusion coefficients for each component in the mixture.
The TEKMC optimization was also carried out independently for each mixture at different pressures,  to obtain the diffusivities.
\begin{table}[t]
    \caption{Diffusion coefficients, solubility and permeability of each species for different binary gas mixtures inside 6F-CMSM.}
    \label{tab:mixture}
    \footnotesize
    \centering
    \begin{tabular}{ccccccc}
    \hline
P (bar) & \multicolumn{2}{c}{$D_i$ $(10^{-6}$ cm$^2$s$^{-1})$} & \multicolumn{2}{c}{$S_i$ $(10^{-2}$ cm$^3_{[STP]}$cm$^{-3}$cmHg$)$} & \multicolumn{2}{c}{$\mathcal{P}_i$ $(10^3$ Barrer$)$} \\
    \hline \hline
\multicolumn{7}{c}{\ce{CO2} : \ce{CH4} $(50:50)$} \\
 & \ce{CO2}  & \ce{CH4} & \ce{CO2} & \ce{CH4} & \ce{CO2}  & \ce{CH4} \\
    \hline
4  & 1.34 $\pm$ 0.12 & 0.78  $\pm$ 0.26 & 39.96 $\pm$ 3.00 & 4.85  $\pm$ 0.76 & 5.34  $\pm$ 0.61  & 0.38 $\pm$ 0.14 \\
6  & 1.65 $\pm$ 0.20 & 0.50  $\pm$ 0.13 & 29.78 $\pm$ 2.22 & 3.59  $\pm$ 0.56 & 4.91  $\pm$ 0.69  & 0.18 $\pm$ 0.06 \\
10 & 3.35 $\pm$ 0.14 & 4.75  $\pm$ 0.83 & 20.09 $\pm$ 1.49 & 2.35  $\pm$ 0.38 & 6.72  $\pm$ 0.58  & 1.12 $\pm$ 0.27 \\
15 & 3.42 $\pm$ 0.44 & 4.24  $\pm$ 0.78 & 14.36 $\pm$ 1.03 & 1.68  $\pm$ 0.26 & 4.91  $\pm$ 0.72  & 0.71 $\pm$ 0.11 \\
20 & 7.02 $\pm$ 0.67 & 13.55 $\pm$ 0.78 & 11.22 $\pm$ 0.78 & 1.27  $\pm$ 0.20 & 7.88  $\pm$ 0.93  & 1.73 $\pm$ 0.28 \\
    \hline
\multicolumn{7}{c}{\ce{CO2} : \ce{CH4} $(10:90)$} \\
 & \ce{CO2}  & \ce{CH4} & \ce{CO2} & \ce{CH4} & \ce{CO2}  & \ce{CH4} \\
    \hline
4  & 2.07 $\pm$ 0.07 & 0.73  $\pm$ 0.26 & 78.28 $\pm$ 28.29 & 8.55 $\pm$ 0.80 & 16.16 $\pm$ 5.87  & 0.62 $\pm$ 0.23 \\
6  & 5.87 $\pm$ 0.35 & 7.25  $\pm$ 0.52 & 61.31 $\pm$ 19.76 & 6.61 $\pm$ 0.56 & 35.99 $\pm$ 11.80 & 4.80 $\pm$ 0.53 \\
10 & 5.4  $\pm$ 0.43 & 9.92  $\pm$ 0.53 & 43.21 $\pm$ 10.66 & 4.62 $\pm$ 0.3 & 23.31  $\pm$ 6.04  & 4.59 $\pm$ 0.39 \\
15 & 2.12 $\pm$ 0.20 & 1.47  $\pm$ 0.13 & 32.34 $\pm$ 7.76 & 3.41  $\pm$ 0.22 & 6.87  $\pm$ 1.77  & 0.50 $\pm$ 0.06 \\
20 & 7.41 $\pm$ 0.36 & 12.90 $\pm$ 0.82 & 26.01 $\pm$ 5.64 & 2.70  $\pm$ 0.16 & 19.27 $\pm$ 4.28  & 3.48 $\pm$ 0.30 \\
    \hline
\multicolumn{7}{c}{\ce{CO2} : \ce{N2} $(20:80)$} \\
 & \ce{CO2}  & \ce{N2}  & \ce{CO2} & \ce{N2} & \ce{CO2}  & \ce{N2} \\
    \hline
4  & 1.50 $\pm$ 0.12 & 3.00  $\pm$ 0.45 & 71.91 $\pm$ 7.66 & 2.03  $\pm$ 0.49 & 10.77 $\pm$ 1.44  & 0.61 $\pm$ 0.17 \\
6  & 2.14 $\pm$ 0.19 & 2.80  $\pm$ 0.16 & 56.13 $\pm$ 5.41 & 1.58  $\pm$ 0.35 & 11.98 $\pm$ 1.56  & 0.44 $\pm$ 0.10 \\
10 & 2.44 $\pm$ 0.16 & 2.11  $\pm$ 0.19 & 39.59 $\pm$ 3.53 & 1.14  $\pm$ 0.23 & 9.65  $\pm$ 1.07  & 0.24 $\pm$ 0.05 \\
15 & 1.44 $\pm$ 0.08 & 2.27  $\pm$ 0.22 & 29.57 $\pm$ 2.47 & 0.86  $\pm$ 0.02 & 4.26  $\pm$ 0.43  & 0.20 $\pm$ 0.04 \\
20 & 2.35 $\pm$ 0.16 & 3.18  $\pm$ 0.22 & 23.81 $\pm$ 2.00 & 0.67  $\pm$ 0.01 & 5.59  $\pm$ 0.60  & 0.21 $\pm$ 0.04 \\
    \hline
\end{tabular}
\end{table}
This data is obtained from $\SI{5}{\micro\second}$ long TEKMC simulations and the diffusivities were computed from the MSD data over $\SI{2.5}{\micro\second}$ to sample the Brownian diffusive regime (Fig.~\ref{fig:binary_exponenet}).
For the binary mixtures, the permeability  $\mathcal{P}_i$ for the $i^{th}$ component is obtained from the corresponding solubility $S_i$.
$S_i$ values are obtained using Eq.~\ref{eq 2.1}, from  binary mixture GCMC simulations reported in our previous work~\cite{DASGUPTA2022121044}. 
The corresponding values of $D_i$, $S_i$ and $\mathcal{P}_i$ for the different gas mixtures are given in Table~\ref{tab:mixture}. 
The variations of the diffusive exponent $\alpha$ for the mixtures are illustrated in Fig.~\ref{fig:binary_exponenet} of the appendix.
We observe that the diffusivities of \ce{CO2}, \ce{CH4}, and \ce{N2} in their binary mixtures deviate significantly from  the pure component values.
$D_{CO_2}$ is greater in majority of the binary mixtures when  compared to diffusivity of pure \ce{CO2} for a given pressure value reflecting the altered energy landscape for \ce{CO2} in the mixture.
The diffusivity of both \ce{CH4} and \ce{N2} are higher than that of \ce{CO2} particularly at higher pressures.
The difference in the mixture diffusivities of the two species is higher for \ce{CH4} than \ce{N2}.
However, the high adsorption of \ce{CO2} onto 6F-CMSM leads to significantly higher $S_{CO_2}$ which in turn results in a greater increase for  $\mathcal{P}_{CO_2}$ when compared with \ce{CH4} and \ce{N2}.
We also note a decrease in the diffusivities for the \ce{CO2} : \ce{CH4} ($10:90$) at a pressure of 15 bar, however the reason for this particular deviation is not completely understood as the MSD data was well within the diffusive regime for both species.

\begin{figure}[t!]
    \centering
    \includegraphics[width=0.9\textwidth]{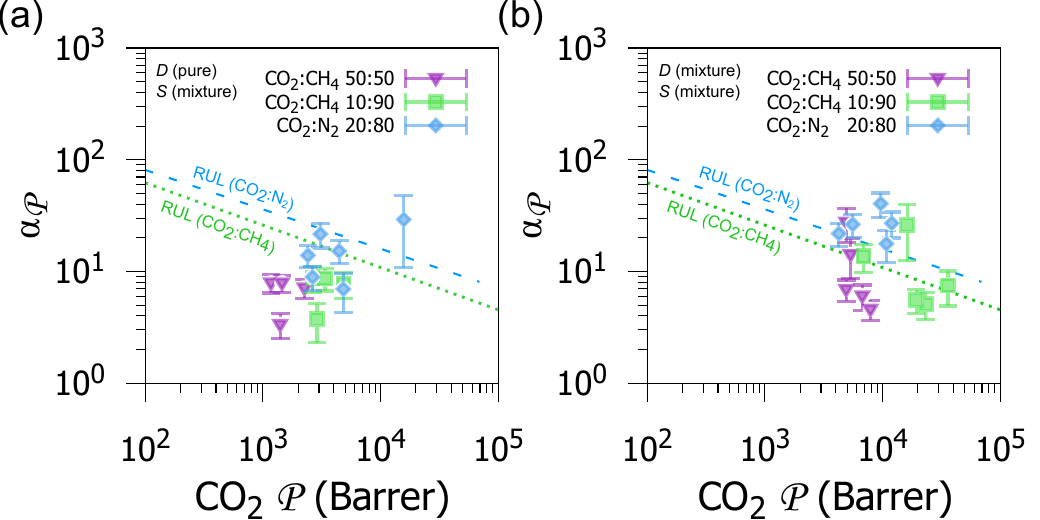}
    \caption{(a) Permselectivity $(\alpha_\mathcal{P})$ calculated using diffusion $(D)$ from pure gas simulations and solubility $(S)$ from binary gas mixtures inside 6F-CMSM.
    (b) $\alpha_\mathcal{P}$ calculated using both $D$ and $S$ from binary gas mixtures.
    $\mathcal{P}$ and $\alpha_\mathcal{P}$ from computed from binary gas simulations are higher than that from pure gas simulations.
    The dashed lines correspond to the Robeson upper limit (RUL) for different gases~\cite{robeson2008upper}.
    $\alpha_{\mathcal{P}}$ obtained from mixture simulations are significantly higher from pure simulation values.
    }
    \label{fig:robeson}
\end{figure}

Fig.~\ref{fig:robeson}a shows the Robeson plots for different mixtures where we have used the solubility mixture data with pure component diffusivity values to obtain the permeabilities.
In Fig.~\ref{fig:robeson}b mixture data (Table~\ref{tab:mixture}) was used for both the solubilities and diffusivities.
The Robeson upper limit for different \ce{CO2}:\ce{CH4} and \ce{CO2}:\ce{N2} are plotted for comparison~\cite{robeson2008upper}.
Upon comparing Fig.~\ref{fig:robeson}a, b, we first note that both $\alpha_{\mathcal{P}}$ and $\mathcal{P}$ increase while using values for $S_i$  and $\mathcal{P}_i$ computed from mixture data.
Using pure component $D_i$ values for the Robeson plot results in a lowering of the corresponding values leading to underpredicting the performance of a given material.
In Fig.~\ref{fig:robeson}a we observe a single point for the \ce{CO2}:\ce{N2} $(20:80)$ mixture that lies above the Robeson limit.
This point corresponds to the lowest pressure of 0.6 bar for which we did not evaluate the corresponding mixture diffusivities.
Interestingly we observe that the data for the \ce{CO2}:\ce{N2} $(20:80)$ mixture lies at or above the Robeson upper limit when mixture diffusivity data is used (Fig.~\ref{fig:robeson}b).
Similar trends are observed for the \ce{CO2}:\ce{CH4} $(10:90)$ mixture as well as the \ce{CO2}:\ce{CH4}  ($50:50$).
The increase in $\mathcal{P}_i$ for \ce{CO2} is primarily due to the increase in the diffusivities of \ce{CO2} in the mixtures compared to the pure gas diffusivities.
The increased permselectivities, $\alpha_{\mathcal{P}}$ are due to a higher diffusivity ratio obtained with mixture diffusivities. 

Our results indicate that using values of single component gas diffusivities can lead to an underestimation of the true permselectivity of a membrane.
Hence obtaining accurate estimates of the diffusivity in both experiments and simulations are needed in order to assess the separation performance of a given material. 
This  will also facilitate a better  understanding of the Robeson upper limits possessed by different membranes synthesized in the laboratory as well as from in silico predictions using MD simulations and machine learning based predictions~\cite{YANG2020631,sanat_sciadv.aaz4301}.
We finally point out that although we have observed these trends for a specific adsorbent, in this 
case the 6F-CSM membrane, a similar analysis would be required to assess the generality of the trends
observed in this study.

%%%%%%%%%%%%%%%%%%%%%%%%%%%%%%%%%%%%%%%%%%%%%%%%%%%%
%% Conclusion
%%%%%%%%%%%%%%%%%%%%%%%%%%%%%%%%%%%%%%%%%%%%%%%%%%%%
\section{Conclusion and outlook}
\label{sec:conclusion}

In this work we obtained the permeability of \ce{CO2}, \ce{CH4} and \ce{N2} gases inside a 6F-CMS membrane for different pressure conditions using multiscale computational techniques, namely all-atom molecular dynamics followed by trajectory extending kinetic Monte Carlo simulations.
The dynamics of adsorbed gas molecules inside the membrane is usually retarded, leading to sub-diffusive mean squared displacements in the timescales of typical all-atom molecular dynamics simulations.
To sample the diffusive dynamics regime, the TEKMC algorithm enables us to extend all-atom MD trajectories to $\SI{5}{\micro\second}$ timescales where the diffusive limit can be reliably sampled.
The constructed transition probability matrices also helps us understand accessible and inaccessible regions of the complex pore networks in the membrane.
In a distinct departure from earlier works we extend the TEKMC simulations to obtain diffusion coefficients of gas mixtures.
To our knowledge this is the first time that mixture diffusivity data has been obtained using this method.  For pure components \ce{CO2}, \ce{N2} and \ce{CH4},  
We report a linear increase in the gas diffusivity confined in  the 6F-CMSM matrix with increasing in pressure.
Conversely both the solubility and permeability of the gases are found to decrease with pressure.

We make a detailed comparison of the permselectivity using both pure component and mixture diffusivity data in the 6F-CMS membrane to assess the extent of deviations observed when pure component diffusivity data are used in lieu of mixture diffusivities.
The implementation of the TEKMC algorithm is available as a package on request.
Gas solubility data are calculated using previous GCMC mixture simulations~\cite{DASGUPTA2022121044}.
\ce{CO2} diffusivities in both \ce{CH4}, and \ce{N2} mixtures in the 6F-CMS membrane were found to be higher than the pure component values.
This leads to a significant increase in both the permselectivity and permeabilities with the use of mixture diffusivities on the Robeson plots, with data lying at or above the Robeson upper limit for the \ce{CO2}:\ce{N2} $(20:80)$ mixtures.
Our analysis indicates that for the \ce{CO2}:\ce{CH4} and  \ce{CO2}:\ce{N2} mixtures investigated in this study, pure component diffusivities  underestimates the membrane performance by varying amounts.
Our findings emphasizes the importance of obtaining accurate mixture diffusivity data while designing membranes for a given separation process. 
Coupled with GCMC and combined MD and TEKMC simulations our study provides a complete in silico framework that can be routinely used to assess the performance of membranes for gas separation processes. 
\update{
We restricted our analysis to the TEKMC study of self-diffusion coefficients in order to determine the gas permeabilities for the Robeson plots to assess separation performance in the polymeric membranes.
The TEKMC analysis can potentially be extended to obtain the  mutual diffusion coefficients, determined from cross correlations between different species in the mixture.~\cite{jamali_jctc}.
Correlations between species in a mixture are a function of the pore size, gas loading and interactions between the gas and the surface. A detailed analysis is needed to fully assess the determination of transport diffusivities while evaluating the gas permeabilities in polymeric membranes.}

\update{\section{Data Availability Statement}
The data underlying this study are available on request.
The code is openly available at \url{https://github.com/PKMLab/tekmc}.
}

\begin{acknowledgement}

The authors thank the Department of Science and Technology, India for funding and providing computational resources.

\end{acknowledgement}

\section{Appendix}
\renewcommand{\thetable}{A\arabic{table}}
\renewcommand{\thefigure}{A\arabic{figure}}
\renewcommand{\thesection}{A\arabic{section}}

\setcounter{table}{0}
\setcounter{figure}{0}
\setcounter{section}{0}

\subsection{\update{Force field parameters}}
\update{
The interactions involving the gas molecules are modeled using the TraPPE force fields~\cite{trappe} and the 6F-CMSM atoms are modelled using modified Dreiding~\cite{roy2020investigations, mayo1990dreiding}.
The values are shown in Table~\ref{table:parameters}.
}
\begin{table}[!ht]
	\centering
 	\caption{\update{Parameters for LJ potential and atomic charges for gas molecules and 6F-CMSM.}}
 	\label{table:parameters}
	\begin{tabular}{c c c c c c}
		\hline & \\[-2ex]
		Molecule & Site &  $\sigma$ (\AA) & $\epsilon$ (K) & q (e) & Bond length (\AA)	\\
	 & \\[-2ex]	\hline
	 	CMS	
	 				& H			& $1.65$	& $7.65$	& $ \quad 0.056$	& $-$	\\	
					& N\_R		& $3.26$	& $38.95$	& $-0.617$    	& $-$	\\
					& C\_3		& $3.47$	& $47.86$	& $ \quad 0.059$	& $-$	\\
					& C\_R\_r	& $3.47$	& $47.86$	& $-0.078$    	& $-$	\\
					& C\_R		& $3.47$	& $47.86$	& $-0.078$	    & $-$	\\
		\hline
		CH$_{4}$	& CH$_{4}$	& 3.73 	& 148.0	& \quad $-$		& $-$			  	\\
					&			&		&		&				& 				  	\\
		CO$_{2}$	& C			& 2.80	& 27.0	& $ \quad 0.70$	& C $-$ O $= 1.16$	\\
					& O			& 3.05	& 79.0	& $-0.35$	 	& 				  	\\
					&			&		&		&				& 				  	\\
		N$_{2}$		& N			& 3.31	& 36.0	& $-0.482$		& N $-$ N $= 1.1$ 	\\
					& COM		& 0.0	& 0.0	& $\quad 0.964$	&				  	\\
		\hline

	\end{tabular}
\end{table}
\subsection{Validation of TEKMC algorithm using bulk water simulations}
\label{sec:validation}
%%%%%%%%%%%%%%%%%%%%%%%%%%%%%%%%%%%%%%%%%%%%%%%%%%%%
% Bulk water
%%%%%%%%%%%%%%%%%%%%%%%%%%%%%%%%%%%%%%%%%%%%%%%%%%%%
We performed a fully atomistic MD simulation of the bulk water system consisting of 800 SPC/E ~\cite{berendsen1987missing} water molecules using LAMMPS.
Initial positions of water molecules were randomly assigned inside the MD box with a constraint that all of them are separated by a threshold distance greater than the bond length of water. 
Periodic boundary conditions were enforced in all directions. 
The bulk water system was energetically minimized to remove bad contacts. 
The minimization comprised of 500 steps of steepest descent followed by 500  steps of conjugate gradient. 
After the energy minimization, the system was heated to \SI{300}{\kelvin} in steps of \SI{30}{\kelvin}, each for 10 ps, using a Langevin thermostat~\cite{schneider1978molecular}.
The system is next simulated in an  NPT ensemble so that the system attains appropriate density at 1 atm pressure. 
During the NPT run of \SI{1}{\nano\second} at \SI{300}{\kelvin} temperature and 1 atm pressure, the density of the system reaches equilibrium by varying the MD box boundaries. 
To maintain a constant pressure of 1 atm, Nosé-Hoover barostat was used with a coupling constant of \SI{2}{\pico\second}. 
Finally, we perform a production run for \SI{1}{\nano\second} in an NVT ensemble. 
Nosé-Hoover thermostat~\cite{nose1984unified, hoover1985canonical} with a heat bath coupling constant of \SI{1}{\pico\second} was used to maintain a constant temperature during the production run. 
The velocity-Verlet scheme was used to do the MD integration with an integration time step of \SI{1}{\femto\second}. 
The OH bond was constrained using the SHAKE~\cite{ryckaert1977numerical} algorithm.
To compute the long-range Coulomb potential, particle-particle particle-mesh Ewald summation method (PPPM)~\cite{hockney2021computer} was used with a tolerance of $10^{-5}$.
The trajectory of water molecules was dumped every \SI{10}{\pico\second}.

The TEKMC algorithm uses trajectories obtained from the production run to extend the MSD to a longer time scale.
Trajectories obtained from the NVT simulation are used by the TEKMC algorithm to extend the MSDs to larger time scales. 
The timestep between the MD frames analyzed by TEKMC is \SI{10}{\pico\second}.
Hence, the timestep between steps of the random walk is \SI{10}{\pico\second}.
From the \SI{1}{\nano\second} NVT simulation of SPC/E water, the transition probability matrix is estimated during the first stage of the TEKMC algorithm. 
We perform 3000 random walks for various grid sizes up to \SI{3}{\nano\second} and tune the grid size to match with the MSD of MD simulation.

Random walks of TEKMC were extended till \SI{100}{\nano\second} for these grid sizes. 
MSD and diffusion coefficient till \SI{0.5}{\nano\second} obtained from the MD simulation are also shown. 
Grid size of \SI{0.28}{\nano\meter} best matches the MD data (having the lowest MSE).
The exponent $\alpha$ is 1.0, indicating that MSD is linear with time and SPC/E water follows Fickian diffusion in this time scale. 
From TEKMC, we obtain diffusion coefficient of SPC/E water to be \update{$2.47 \pm \SI{0.15e-5}{\centi\meter\squared\per\second}$}, in good agreement to the literature value of $\SI{2.75e-5}{\centi\meter\squared\per\second}$~\cite{doi:10.1021/jp003020w}.
\update{
The diffusion coefficient of bulk water obtained from simulations is subject to finite size effects based on the size of the box and number of molecules~\cite{ioannis_2019}.
The main goal of our validation was to reproduce the MD obtained MSD by tuning $d_{grid}$ during TEKMC.
Once validated, our main focus is on the gas diffusivity inside the ultramicropores.
Diffusion in the polymeric membrane is related to the inherent pore size distribution in the membrane which could be influenced by finite size effects.
We point out that the polymeric membranes used in this study have been prepared by using extensive compression and decompression cycles which has shown to reproduce the experimental pore size distribution and adsorption isotherm for this system~\cite{DASGUPTA2022121044}.
The finite size corrected diffusivities may also be obtained by a suitable further tuning of $d_{grid}$.
}
The self-diffusion coefficient of pure water has been measured to be  $\SI{2.3e-5}{\centi\meter\squared\per\second}$ at \SI{298}{\kelvin} from experimental techniques like  diaphragm-cell technique\cite{mills1973self} or the pulsed-gradient spin echo (PGSE) NMR method\cite{price1999self}.

\newpage
\subsection{Probability distribution of gases used for performing TEKMC}
\label{sec:pure}

\begin{figure}[!h]
    \centering
    \includegraphics[height=9cm]{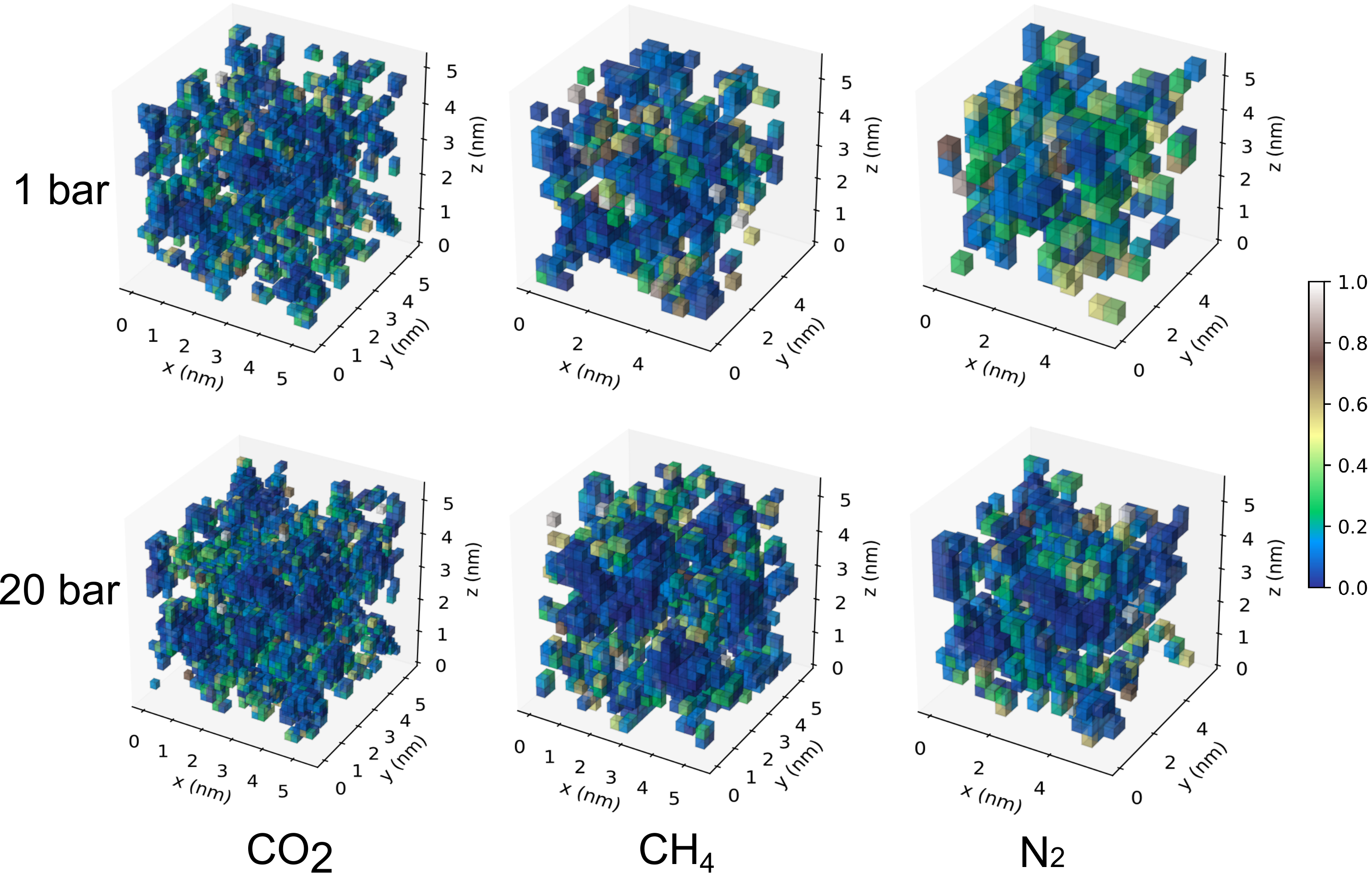}
    \caption{
    Probability of survival in a voxel for pure gas systems at two pressure values.
    The size of the voxel corresponds to the optimum $d_{grid}$ per system.
    }
    \label{fig:probability}
\end{figure}

\begin{figure}[!h]
    \centering
    \includegraphics[height=9cm]{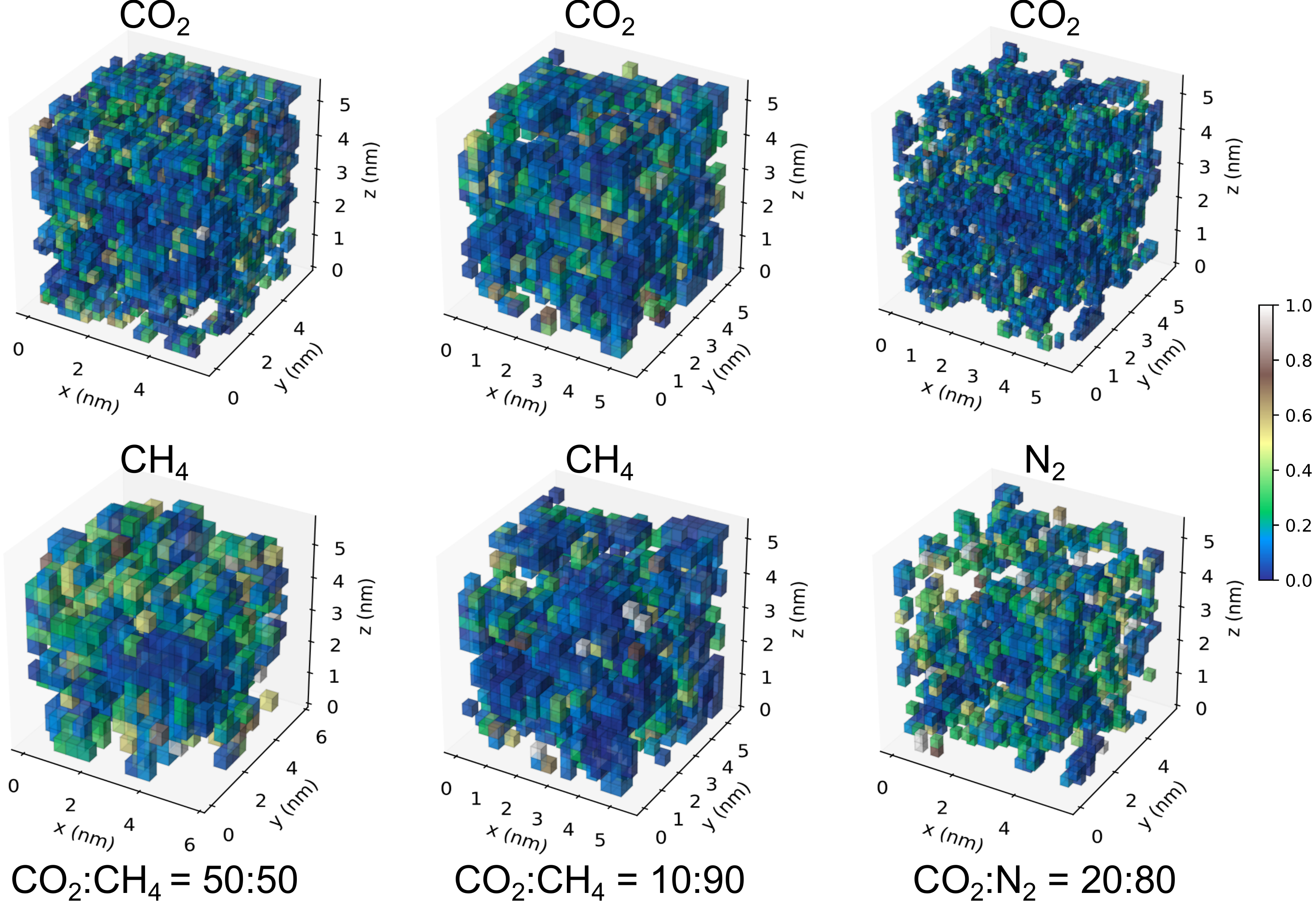}
    \caption{
    Probability of survival in a voxel for different binary gas mixture systems at $\SI{20}{bar}$ pressure.
    }
    \label{fig:mix_probability}
\end{figure}

\newpage
\subsection{Binary component diffusion inside 6F-CMSM}
\label{sec:binary}

\begin{figure}[h!]
    \centering
    \includegraphics[height = 9cm]{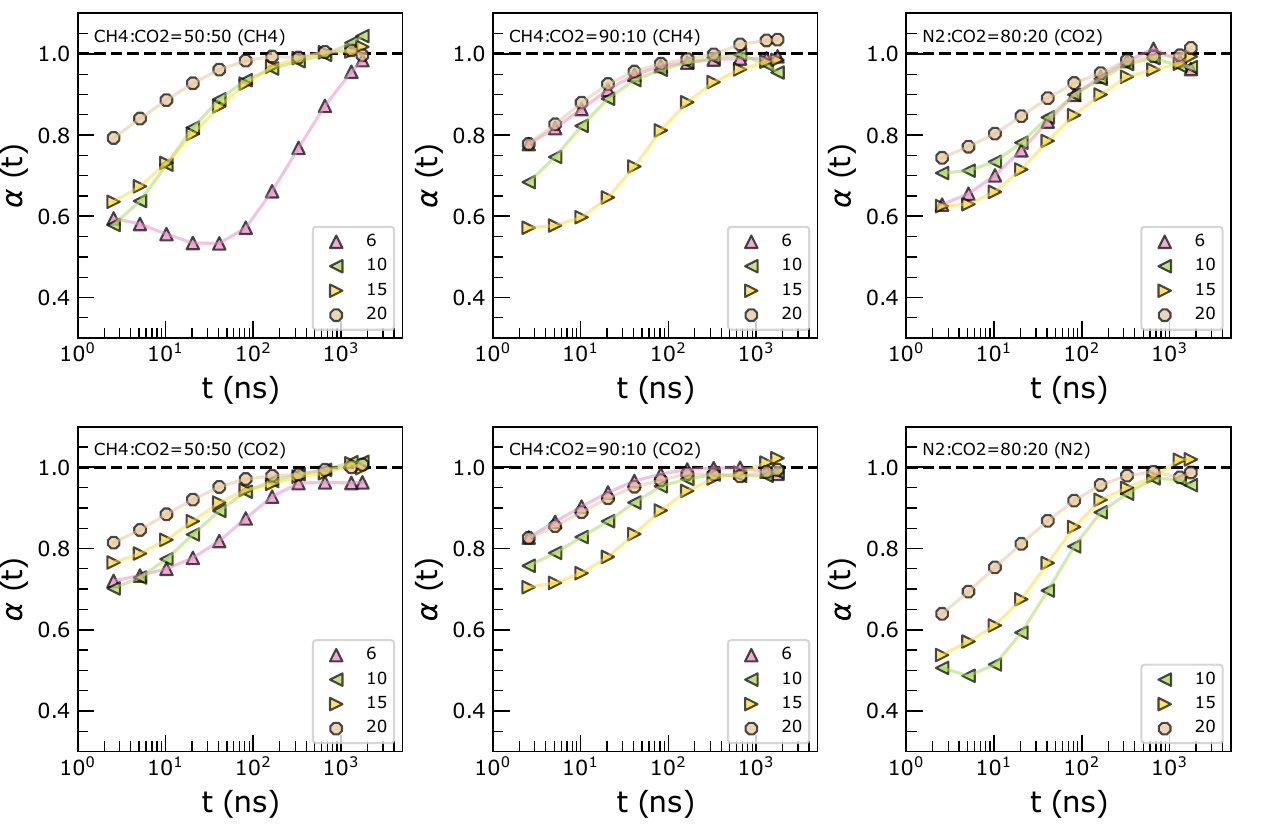}
    \caption{
    Time series of MSD exponent $(\alpha)$ for the different binary mixture systems studied.
    The legend shows different pressure, P (bars).
    }
    \label{fig:binary_exponenet}
\end{figure}

We perform MD simulations of binary gas mixtures inside 6F-CMSM corresponding to different pressure conditions.
The obtained trajectories are then extended till $\SI{5}{\micro\second}$ to obtain mixture diffusion coefficients of the different gases.
We find the diffusivities are significantly different in the mixture compared with the pure gases shown in the main accompanying text.
The diffusivity of \ce{CO2} is comparable to \ce{CH4} or \ce{N2} only for low pressures.
The diffusivity is also affected by the molar ratio of the mixtures.

\bibliography{reference}

\end{document}